\documentclass[11pt]{article}
\usepackage{graphicx}
\usepackage{epsfig}

\newcommand{\eps}{\varepsilon}
\newcommand{\pfrac}[2]{\left(\frac{#1}{#2}\right)} 

\begin{document}
\begin{flushright}
MZ-TH/04-03\\
hep-ph/0403122\\
March 2004\\
\end{flushright}
\vspace{0.95cm}

\begin{center}
\renewcommand{\thefootnote}{\fnsymbol{footnote}}
{\Large\bf Laurent series expansion of sunrise-type diagrams\\[12pt]
using configuration space techniques\footnote[1]{Partially supported by 
RFBR grants \# 02-01-601, 03-02-17177.}}
\renewcommand{\thefootnote}{\arabic{footnote}}
\vspace{1truecm}

{\large \bf S.~Groote$^{1,2}$, J.G.~K\"orner$^1$ and
A.A.~Pivovarov$^{1,3}$}\\[.4truecm] 
$^1$ Institut f\"ur Physik, Johannes-Gutenberg-Universit\"at,\\
Staudinger Weg 7, D-55099 Mainz, Germany\\[.3truecm]
$^2$ F\"u\"usika-Keemiateaduskond, Tartu \"Ulikool,
T\"ahe 4, EE-51010 Tartu, Estonia\\[.3truecm] 
$^3$ Institute for Nuclear Research of the\\
Russian Academy of Sciences, Moscow 117312, Russia
\end{center}

\addtocounter{footnote}{3}

\begin{abstract}
We show that configuration space techniques can be used to efficiently
calculate the complete Laurent series $\eps$-expansion of sunrise-type
diagrams to any loop order in $D$-dimensional space-time for any external
momentum and for arbitrary mass configurations. For negative powers of
$\eps$ the results are obtained in analytical form. For positive powers of
$\eps$ including the finite $\eps^0$ contribution the result is obtained
numerically in terms of low-dimensional integrals. We present general
features of the calculation and provide exemplary results up to five
loop order which are compared to available results in the literature.
\end{abstract}

\newpage

\section{Introduction}
The computation of multi-dimensional integrals corresponding to multi-loop
Feynman diagrams is a necessity for high precision calculation in quantum
field theory since perturbation theory remains the main tool of most
theoretical analysis' within the Standard Model and
beyond~\cite{SMrev,Hollik:1999uy,Blumlein:qs,threeloops}. An important
ingredient of the algebraic approach to the evaluation of Feynman diagrams
is the integration-by-parts technique which allows one to derive and analyze
recurrence relations for the sets of relevant multi-loop
integrals~\cite{ibyparts}. Using the integration-by-parts technique, many
diagrams are reduced to simplified subsets of master
integrals (see e.g.~\cite{Broadhurst:1991fi,Baikov:1996rk,Chetyrkin:1996cf,
Laporta:1996mq,Tarasov:1997kx,Gehrmann:1999as,Melnikov:2000zc,Laporta:2001dd,
Grozin:2002zb}).

An important subset of master integrals is represented by diagrams of the
sunrise topology, the so-called sunrise-type diagrams (also called
sunset-type, water-melon, basketball, or banana
diagrams)~\cite{Meijer,ellRef,Mendels}.
Diagrams of the sunrise topology have been studied quite intensively in the
past and many of their properties have been known for quite some 
time~\cite{Broadhurst:1991fi,Avdeev:1995eu,Groote:1999cn,Laporta:2001dd,
Schroder:2002re}. Physical applications are also numerous as worked out in 
e.g.~\cite{Larin:1986yt,Sakai:1999qm,Groote:2001vr,bar,Narison:1994zt}. This
concerns in particular finite temperature calculations using effective 
potentials~\cite{effpot,Gross:1980br,Andersen:2000zn,Nishikawa:2003js,
Rajantie:1996cw}. There are also applications in nuclear and solid state
physics where methods of quantum field theory are used for the construction of
low-energy effective theories~\cite{nuclth}. The importance of the sunrise
topology in physical applications alone justifies the strong interest in
sunrise-type diagrams~\cite{Witten:1979kh,Kajantie:2003ax,Post:1997dk,
Post:1996gg,Gasser:1998qt,Delbourgo:2003zi,Bashir:2001ad,Ligterink:1999mu}. At
the same time this topology is a good laboratory for checking the efficiency
of new methods of multi-loop calculations~\cite{Davydychev:1999ic,Mendels:qe,
Caffo:2002ch}. The relevant master integrals should be calculated as precisely
as possible within dimensional regularization, the results of which should be
expressed in terms of the complete Laurent series expansion in the dimensional
parameter $\eps=(D-4)/2$ including also positive powers of
$\eps$~\cite{Bierenbaum:2003ud}. Higher order terms in the $\epsilon$
expansion are needed if the sunrise-type diagram is inserted into a divergent
diagram or when one is using the integration-by-parts recurrence relation
which can generate inverse powers of $\epsilon$. For example, recently the
Laurent series expansion for four-loop master integrals has been found using
numerical methods~\cite{Laporta:2002pg}. The numerical evaluation of Feynman
diagrams has become a valuable tool in higher order loop calculations because
the extreme complexity of these calculations often precludes an analytical
approach and thus the numerical approach is the only possibility to obtain
physical results~\cite{Passarino:2001wv}.

In this paper we develop and describe a configuration space method which
allows one to efficiently evaluate the coefficients of the Laurent series
expansion in the dimensional parameter $\eps$ for sunrise-type
diagrams~\cite{Groote:1998ic}. The method is simple to use and it works
well for arbitrary mass configurations, arbitrary values of the external
momentum and any number of loops. 

\begin{figure}[ht]\begin{center}
\epsfig{figure=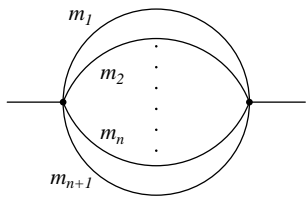, scale=0.3}
\caption{\label{fig1}$n$-loop diagram with $(n+1)$ lines of the sunrise-type 
topology}
\end{center}\end{figure}

In configuration space, the correlator function described by a sunrise-type 
diagram with $(n+1)$ lines (corresponding to $n$ loops, cf.\ Fig.~\ref{fig1})
is given by the product of propagators $D(x,m)$,
\begin{equation}\label{eqn01}
\Pi_n(x)=\prod_{i=1}^{n+1}D(x,m_i)
\end{equation}
and/or their derivatives if necessary (for details see
Refs.~\cite{Groote:1998ic,Groote:1998wy,Groote:2000kz}). The propagator
$D(x,m)$ of a massive particle with the mass parameter $m$ in $D$-dimensional
(Euclidean) space-time is given by the momentum space integral which can be
evaluated in terms of Bessel functions,
\begin{equation}\label{eqn02}
  D(x,m)=\frac1{(2\pi)^D}\int\frac{e^{-i(p\cdot x)}d^Dp}{p^2+m^2}
  =\frac{(mx)^\lambda K_\lambda(mx)}{(2\pi)^{\lambda+1}x^{2\lambda}}
\end{equation}
where we write $D=2\lambda+2$. Here $K_\lambda(z)$ is a modified Bessel
function of the second kind (see e.g.~\cite{Watson}). In the zero mass case
the propagator simplifies to
$D(x,0)=\Gamma(\lambda)/4\pi^{\lambda+1}x^{2\lambda}$.
It is obvious that the correlator function $\Pi(x)$ in configuration space
contains no integration at all. In this respect it is a kind of analogue to a 
tree diagram in momentum space. A calculation of a sunrise-type diagram is
necessary only if one wants to calculate the diagram in momentum space. This
requires a Fourier transformation,
\begin{equation}\label{eqn05}
\tilde\Pi_n(p)=\int\Pi_n(x)e^{i(p\cdot x)}d^Dx.
\end{equation}
Note that the required integrals are basically scalar which makes the angular 
integration in Eq.~(\ref{eqn05}) simple in $D$-dimensional space-time. One
obtains one-dimensional integrals
\begin{equation}\label{eqn07}
\tilde\Pi_n(p)=2\pi^{\lambda+1}\int_0^\infty\pfrac{px}2^{-\lambda}
  J_\lambda(px)\Pi_n(x)x^{2\lambda+1}dx,
\end{equation}
where $p=|p|$ and $x=|x|$ are the absolute values of $p^\mu$ and $x^\mu$,
respecticely, and $J_\lambda(z)$ is the Bessel function of the first kind.
For integrals with additional tensor structure the plane wave function
$e^{i(p\cdot x)}$ occurring in the Fourier integral in Eq.~(\ref{eqn05}) has
to be expanded in a series of Gegenbauer polynomials
$C_j^\lambda(w)$~\cite{Chetyrkin:pr,Terrano:1980af}. Because this expansion
does not change the principal structure of the expressions, the representation
given by Eq.~(\ref{eqn07}) is quite universal and will be used in the
following.

The paper is organized as follows. In Sec.~2 we explain the treatment of the
singular part of the integral while the calculation of the nonsingular part is
dealed with in Sec.~3. After a striking but quite simple example we present
four- and five-loop calculations of the bubble diagram in Sec.~4. Finally, in
Sec.~5 we give recipes to treat so-called irreducible numerator factors in
sunrise diagrams. In Sec.~6 we present our conclusions.  

\section{$\eps$-expansion of the singular part}
The ultraviolet (UV) singularities that appear in the correlator function
$\tilde\Pi(p)$ are related to the small $x$ behaviour of the integrand in
Eq.~(\ref{eqn07}). These singularities can be subtracted using the standard
$R$-operation~\cite{Bogoliubov}. The simplest way to do this is to expand the
relevant Bessel functions at small $x$ and then to integrate the resulting
integrand in $D=4-2\eps$ dimensions. However, in order to avoid infrared
singularities coming from the large $x$-region of integration, some parts of
the unexpanded Bessel functions should be retained. From a technical point of
view one or two Bessel functions in the integrands can be kept unexpanded to
provide the necessary cutoff. This is possible because the integrals
containing one or two Bessel functions in the integrand can be done
analytically. Indeed, for any $\mu,\nu$ one has~\cite{Prudnikov,Gradshteyn}
\begin{eqnarray}\label{eqn11}
\int_0^\infty x^{\mu-1} K_\nu(mx)dx&=&\frac{2^{\mu-2}}{m^\mu}
  \Gamma\pfrac{\mu+\nu}2\Gamma\pfrac{\mu-\nu}2,\nonumber\\
\int_0^\infty x^{\mu-1}K_\nu(mx)K_\nu(mx)dx
  &=&\frac{2^{\mu-3}}{m^\mu\Gamma(\mu)}\Gamma\left(\frac\mu2+\nu\right)
  \Gamma\pfrac\mu2\Gamma\pfrac\mu2\Gamma\left(\frac\mu2-\nu\right).
\end{eqnarray}
However, because one or two of the Bessel functions are kept unexpanded, this
method breaks the natural symmetry between the masses of different lines. An
alternative recipe is to introduce a damping factor into the integrand with
an adequate power of $n$ such as e.g.\ 
\begin{equation}
h(x)=e^{-\mu^2 x^2}\sum_{k=0}^n\frac{(\mu x)^{2k}}{k!}=e^{-\mu^2x^2}
  \left\{1+\mu^2x^2+\frac12\mu^4x^4+\frac16\mu^6x^6+\ldots\right\}.
\end{equation}
One then expands all massive propagators for small $x$ keeping only the
singular terms of the integrand. For the extraction of the UV poles this is
the most convenient way to proceed. At small $x$ the damping factor behaves
as $h(x)=1+O(x^{2n})$ which does not change the structure of the singularities
of the integrand at small $x$ if $n$ is sufficiently large (depending on the
number of lines and the dimension of space-time). Still another possibility is
to use a hard cutoff in configuration space by introducing a cutoff in the
integration at some given $x_0$ and to extract the poles from the integral
over the finite interval (see e.g.\ Ref.~\cite{Braaten:1995cm}). Which of the
three methods is the most suitable depends on the problem at hand. We found
that the damping factor method is most convenient for the calculation of the
pole parts. In the equal mass case it is advantageous to use the Bessel
function method for the calculation of the nonsingular parts of the
$\eps$-expansions since one generates more compact expressions for the
necessary subsequent numerical evaluation. 
 
When one performs a series expansion of the integrand near the origin, one
easily obtains the singular parts for any sunrise-type diagram for any mass
configuration. As an example we consider the genuine two-loop sunrise diagram
with three different loop masses, 
\begin{equation}
\Pi_2(x)=D(x,m_1)D(x,m_2)D(x,m_3)
\end{equation}
for $D=4-2\eps$ space-time dimensions (the index ``$2$'' in $\Pi_2(x)$ stands
for the two-loop case). In this case one needs two UV counterterms for the
renormalization of the divergent integral,
\begin{equation}\label{renormCprac}
\Pi_2^{\rm ren}(x)=\Pi_2^{\rm reg}(x)-C_0\delta(x)-C_2\partial^2\delta(x)
\end{equation}
with $\partial^2=\partial^\mu\partial_\mu$ which will lead to a factor
$-p^2$ in momentum space. The counter terms read
\begin{eqnarray}
C_0&=&{\cal N}_2\int\Pi_2(x)d^Dx
  \ =\ {\cal N}_2\int D(x,m_1)D(x,m_2)D(x,m_3)d^Dx,\nonumber\\
C_2&=&\frac{{\cal N}_2}{2D}\int\Pi_2(x)x^2d^Dx 
  \ =\ \frac{{\cal N}_2}{2D}\int D(x,m_1)D(x,m_2)D(x,m_3)x^2d^Dx.
\end{eqnarray}
It is clear that the program of UV renormalization requires the calculation
of vacuum bubble diagrams.

Note that we use a special normalization convention for the integration
measure. This is the usual integration measure for massive vacuum integrals
that considerably simplifies the expressions for the counterterms (pole parts)
by removing some awkward terms containing the Euler constant $\gamma$ or
$\pi^2$. The normalization factor is given by 
\begin{equation}\label{normfactorN}
{\cal N}_n=\pfrac{(4\pi)^{2-\eps}}{\Gamma(1+\eps)}^n
\end{equation}
for the $n$-loop (or $(n+1)$-line) sunrise-type diagram.

Starting with the singular pieces for the genuine two-loop sunrise diagram
with equal masses $m$, we obtain 
\begin{equation}
C_0=m^2\left\{-\frac3{2\eps^2}-\frac9{2\eps}\right\}+O(\eps^0),
\qquad C_2=\frac1{\eps}+O(\eps^0)
\end{equation}
where the renormalization scale $\mu$ is set to $\mu=m$. After the counterterms
have been determined, external lines carrying external momenta can be added
because the pole parts of the vacuum bubbles and the sunrise-type diagrams are
identical. The generalization of this example to $n$-loop sunrise-type
diagrams with any mass configuration is straightforward. In the main text we
list a few sample results for equal masses, again setting the renormalization
scale $\mu$ to $\mu=m$. This includes sunrise-type diagrams with vanishing
outer momenta called bubble diagrams. We compute the four-loop bubble diagram
also calculated by Laporta~\cite{Laporta:2002pg} whose result we reproduce. In
the sample results below the normalization factor ${\cal N}_n$ is included
according to the number of lines/loops that makes the definition of
$\tilde\Pi_2(p^2)$ a bit different from that used in
Eqs.~(\ref{eqn05},\ref{eqn07}). The sample results are 
\begin{eqnarray}\label{poleparts}
\tilde\Pi_1(p^2)&=&\frac1\eps+O(\eps^0),\nonumber\\
\tilde\Pi_2(p^2)&=&m^2\left\{-\frac3{2\eps^2}-\frac9{2\eps}
  \right\}-\frac{p^2}{4\eps}+O(\eps^0),\nonumber\\
\tilde\Pi_3(p^2=0)&=&m^4\left\{\frac2{\eps^3}+\frac{23}{3\eps^2}
  +\frac{35}{2\eps}\right\}+O(\eps^0),\nonumber\\
\tilde\Pi_3(p^2=-m^2)&=&m^4\left\{\frac2{\eps^3}+\frac{22}{3\eps^2}
  +\frac{577}{36\eps}\right\}+O(\eps^0),\nonumber\\
\tilde\Pi_4(p^2=0)&=&m^6\left\{-\frac5{2\eps^4}-\frac{35}{3\eps^3}
  -\frac{4565}{144\eps^2}-\frac{58345}{864\eps}\right\}+O(\eps^0),\nonumber\\
\tilde\Pi_4(p^2=-m^2)&=&m^6\left\{-\frac5{2\eps^4}-\frac{45}{4\eps^3}
  -\frac{4255}{144\eps^2}-\frac{106147}{1728\eps}\right\}+O(\eps^0),\nonumber\\
\tilde\Pi_5(p^2=0)&=&m^8\left\{\frac3{\eps^5}+\frac{33}{2\eps^4}
  +\frac{1247}{24\eps^3}+\frac{180967}{1440\eps^2}+\frac{898517}{3456\eps}
  \right\}+O(\eps^0),\nonumber\\
\tilde\Pi_5(p^2=-m^2)&=&m^8\left\{\frac3{\eps^5}+\frac{16}{\eps^4}
  +\frac{49}{\eps^3}+\frac{6967}{60\eps^2}+\frac{1706063}{7200\eps}
  \right\}+O(\eps^0),\nonumber\\
\tilde\Pi_6(p^2=0)&=&m^{10}\left\{-\frac7{2\eps^6}-\frac{133}{6\eps^5}
  -\frac{238}{3\eps^4}-\frac{77329}{360\eps^3}-\frac{21221921}{43200\eps^2}
  -\frac{2596372387}{2592000\eps}\right\}+O(\eps^0).\qquad
\end{eqnarray}
Note the dependence on the external momentum $p^2$. This dependence has
its origin in the derivatives appearing e.g.\ in Eq.~(\ref{renormCprac}).
The coefficient of the leading singularity in $\eps$ is independent of $p^2$. 
In Appendix~A we list results for unequal mass configurations up to
four-loop order. When setting all masses equal the results of the general
mass case can be seen to agree with the above equal mass results.

After having determined the singular parts of the Laurent series expansion
using the damping factor method what remains to be done is to calculate the 
coefficients of the positive powers of $\eps$ in the $\eps$-expansion
including the finite $\eps^0$ term. In order to determine the nonsingular
parts one needs to resort to the Bessel function method. What is technically
needed is to develop a procedure for the $\eps$-expansion of Bessel functions.
This will be the subject of the next section.

Starting with the next section we discuss only bubble diagrams which have
a simpler structure and allow for a comparison with results in the literature
(e.g.\ with Ref.~\cite{Laporta:2002pg}). It is clear that nonsingular parts
can also be calculated for sunrise-type diagrams with $p^2\ne 0$ with our
methods. For $p^2>\sum_im_i^2$ the calculation gives rise to absorptive parts
represented by the spectral density (cf.\
Refs.~\cite{Groote:1998ic,Groote:1998wy,Groote:1999cx}) which
will not be discussed any further in this paper.

\section{$\eps$-expansion of the nonsingular part}
The nonsingular parts in the $\eps$-expansion of sunrise-type diagrams can
also easily be calculated with the help of configuration space techniques.
However, in contrast to the singular coefficients calculated in the last
section, in the general case the computation of the nonsingular coefficients
requires a numerical evaluation. A technical problem which appears in the
computation of higher order terms of the $\eps$-expansion is the necessity to
expand the Bessel functions in their indices. For the first order in $\eps$
the relevant corrections are known and can again be re-expressed in terms of
Bessel functions. Using~\cite{Gradshteyn}
\begin{equation}\label{eexpex}
\left[\frac{\partial K_\nu(z)}{\partial\nu}\right]_{\nu=\pm n}
  =\pm\frac12n!\sum_{k=0}^{n-1}\left(\frac z2\right)^{k-n}
  \frac{K_k(z)}{k!(n-k)},\qquad n\in\{0,1,\ldots\,\}
\end{equation}
for the derivative of the Bessel function $K_\nu(z)$ with respect to its index
near integer values of this index, we obtain the series expansion
\begin{eqnarray}
K_{-\eps}(x)&=&K_0(x)+O(\eps^2),\nonumber\\[7pt]
K_{1-\eps}(x)&=&K_1(x)-\frac\eps xK_0(x)+O(\eps^2),\nonumber\\
K_{2-\eps}(x)&=&K_2(x)-\frac{2\eps}xK_1(x)-\frac{2\eps}{x^2}K_0(x)+O(\eps^2).
\end{eqnarray}
for the first few Bessel functions with non-integer indices (for details cf.\
Ref.~\cite{Groote:1999cx}). Note that the first two results suffice to find
the expansion for the Bessel function with any integer index due to the
recurrence relations that connect three Bessel functions with consecutive
indices. To the best of our knowledge one has to proceed numerically for the
higher order terms. At least we were not able to find a general procedure for
the analytical evaluation of the finite parts of the $\eps$-expansion. Some
analytical results can be found in~\cite{Davydychev:2000na,Suzuki:2004ue}.

A convenient starting point for the $\eps$-expansion is the integral
representation
\begin{equation}
K_\nu(z)=\int_0^\infty e^{-z\cosh t}\cosh(\nu t)dt 
\end{equation}
Then the expansion for $K_{-\eps}(z)$ reads
\begin{equation}
K_{-\eps}(z)=\int_0^\infty e^{-z\cosh t}\cosh(-\eps t)dt 
=\sum_{n=0}^\infty\frac{\eps^{2n}}{(2n)!}f_{2n}(z)
\end{equation}
where
\begin{equation}\label{fndef}
f_k(z)=\int_0^\infty t^ke^{-z\cosh t}dt.
\end{equation}
The family of functions $f_k(z)$ is rather close to the original set of
Bessel functions $K_\nu(z)$ and can easily be studied both analytically
and numerically. The limits at $z\to 0$ and $z\to\infty$ are known
analytically and are simple. They allow for an efficient interpolation for
intermediate values of the argument. 

For the expansion of the second basic function $K_{1-\eps}(z)$ we write
\begin{equation}
K_{1-\eps}(z)=\int_0^\infty e^{-z\cosh t}\cosh(t-\eps t)dt
  =\sum_{n=0}^\infty\frac{\eps^{2n}}{(2n)!}a_{2n}(z)
  -\sum_{n=0}^\infty\frac{\eps^{2n+1}}{(2n+1)!}b_{2n+1}(z)
\end{equation}
with 
\begin{equation}
a_k(z)=\int_0^\infty t^ke^{-z\cosh t}\cosh t\,dt,\qquad
b_k(z)=\int_0^\infty t^ke^{-z\cosh t}\sinh t\,dt.
\end{equation}
Integration by parts and parameter derivatives can be used to further reduce
integrals containing any power of the hyperbolic functions $\sinh(t)$ and
$\cosh(t)$ in $f_k(z)$. The functions $f_k(z)$ allow one to calculate the
higher order coefficients in the $\eps$-expansion. Therefore, the two
functions $a_k(z)$ and $b_k(z)$ are again related to the functions $f_k(z)$,
\begin{equation}
a_k(z)=-\frac{d}{dz}f_k(z),\qquad
b_k(z)=\frac{k}{z}f_{k-1}(z).
\end{equation}
Using these relations, we obtain
\begin{equation}
K_{1-\eps}(z)=-\sum_{n=0}^\infty\frac{\eps^{2n}}{(2n)!}
  \left(\frac{d}{dz}+\frac\eps z\right)f_{2n}(z).
\end{equation}
Due to recurrence relations in the index $\nu$ for the family $K_\nu(z)$ these
two formulas suffice to calculate the Bessel function $K_\nu(z)$ for
any $\nu$. 

\begin{figure}[ht]\begin{center}
\epsfig{figure=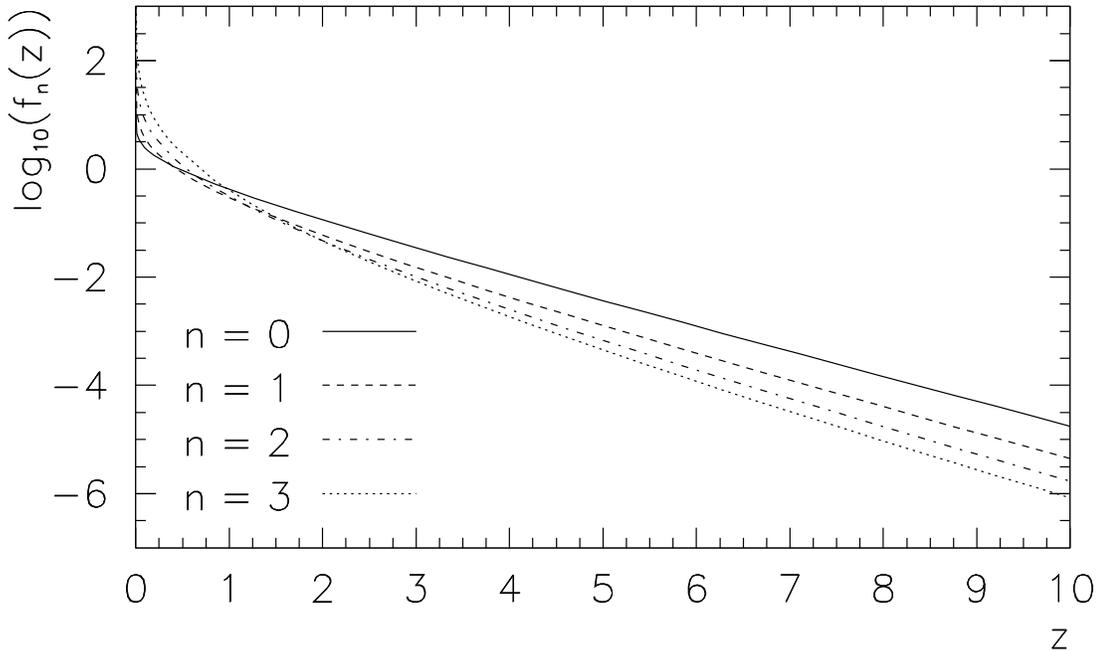, scale=0.9}
\caption{\label{fig2}Comparison of functions $f_n(z)$ for different values of
  $n$, plotted on a logarithmic scale}
\end{center}\end{figure}

The functions $f_k(z)$ satisfy the differential equation
\begin{equation}
\left(\frac{d^2}{dz^2}+\frac1z\frac{d}{dz}-1\right)f_k(z)
  =\frac{k(k-1)}{z^2}f_{k-2}(z)
\end{equation}
related to the Bessel differential equation. The small $z$ behaviour
\begin{equation}
f_k(z)=\frac1{k+1}\ln^{k+1}\pfrac1z
\left(1+O\left(\frac1{\ln(z)}\right)\right)
\end{equation}
can be found by using yet another representation for the function $f_k(z)$,
obtained from Eq.~(\ref{fndef}) by the substitution $u=\cosh t$,
\begin{equation}
f_k(z)=\int_1^\infty\frac{e^{-zu}}{\sqrt{u^2-1}}\ln^k(u+\sqrt{u^2-1})du
\end{equation}
or directly from the behaviour of the function $K_\nu(x)$ at small $x$,
\begin{equation}
K_{-\eps}(x)=\frac1\eps\sinh\left(\eps\ln(1/x)\right).
\end{equation}
In Fig.~\ref{fig2} the functions $f_n(z)$ are compared with $K_0(z)$ for
various values of $n$. Note that $f_0(z)=K_0(z)$ and $a_0(z)=-f'_0(z)=K_1(z)$.
All curves are very smooth and their functional behaviour is similar to that
of the original Bessel function. 

While in four-dimensional space-time the massive propagator contains the
Bessel function $K_{1-\eps}(z)$, for $D=2$ dimensional space-time the basic
function is the Bessel function $K_{-\eps}(z)$, since the propagator reads
\begin{equation}
D(x,m)|_{D=2}=\frac1{2\pi}K_0(mx).
\end{equation}
As an example for the numerical calculation of the $\eps$-expansion using 
Bessel functions we consider a toy model integral related to the one-loop case
in two dimensions. We select this example because the integral is finite and
analytically known, so that we can compare our numerical calculation
with the exact answer. Using Eq.~(\ref{eqn11}), we obtain
\begin{equation}
2\int K_{-\eps}(x)K_{-\eps}(x)x\,dx=\Gamma(1+\eps)\Gamma(1-\eps).
\end{equation}
The expansion in $\eps$ is given by
\begin{equation}
\Gamma(1-\eps)\Gamma(1+\eps)=\frac{\pi\eps}{\sin(\pi\eps)}
  =1+\frac{\pi^2\eps^2}6+\frac{7\pi^4\eps^4}{360}+O(\eps^6)
\end{equation}
On the other hand, we can use the expansion
\begin{equation}
K_{-\eps}(x)=f_0(x)+\frac{\eps^2}2f_2(x)+\frac{\eps^4}{24}f_4(x)
\end{equation}
to rewrite the integral in the form 
\begin{eqnarray}
2\int_0^\infty K_{-\eps}(x)K_{-\eps}(x)x\,dx&=&2\int_0^\infty f_0(x)^2x\,dx
  +2\eps^2\int_0^\infty f_0(x)f_2(x)x\,dx\,+\nonumber\\&&
  +\frac{\eps^4}6\int_0^\infty f_0(x)f_4(x)x\,dx
  +\frac{\eps^4}2\int_0^\infty f_2(x)^2x\,dx+O(\eps^6).\qquad
\end{eqnarray}
Using the explicit expressions for the functions $f_k$ we checked by
numerical integration that the identities
\begin{eqnarray}
2\int_0^\infty f_0(x)^2x\,dx&=&1,\nonumber\\
2\int_0^\infty f_0(x)f_2(x)x\,dx&=&\frac{\pi^2}6,\nonumber\\
\frac16\int_0^\infty(f_0(x)f_4(x)+3f_2(x)^2)x\,dx&=&\frac{7\pi^4}{360}
\end{eqnarray}
are valid numerically with very high degree of accuracy. We have implemented
our algorithm for the $\eps$-expansion of sunrise-type diagrams as a simple
code in Wolfram's MATHEMATICA system for symbolic manipulations and checked
its work-ability and efficiency.

\section{Examples}
In this section we compute some further examples using our techniques. First
we check on a recent result obtained by Laporta for sunrise-type four-loop
bubbles~\cite{Laporta:2002pg}. In Ref.~\cite{Laporta:2002pg} the difference
equation method is used to numerically obtain the coefficients of the Laurent
series expansion of all four-loop bubble master integrals. Our results for the
sunrise-type topology derived by using configuration space techniques provide
an independent check for the results in Ref.~\cite{Laporta:2002pg} where
momentum space techniques were used to calculate the whole set of four-loop
bubble master integrals. This check may help in establishing further
confidence in the results of Laporta. As a new application we present the
five-loop result for the scalar bubble topology.

\subsection{Four-loop vacuum bubble}
The four-loop quantity of interest ($V_1$ in the notation
of~\cite{Laporta:2002pg}) is the master integral
\begin{equation}
\tilde\Pi_4(p^2=0)={\cal N}_4\int D(x,m)^5d^Dx
\end{equation}
where the normalization factor ${\cal N}_4$ is defined in
Eq.~(\ref{normfactorN}). The Laurent series expansion of the four-loop master
integral has been calculated numerically by the difference equation method 
with high precision in~\cite{Laporta:2002pg}. In Sec.~2 we explained how to
obtain the coefficients of its singular part. In fact, we wrote down explicit
results for the singular part of the four-loop master integral. In this
section we shall compute the finite part and the first three coefficient
functions of its $\eps$-expansion numerically using configuration space
techniques.

The general idea is really quite straightforward. As explained before, UV
divergences reveal themselves in configuration space as singularities of the
integrand at small $x$. We subtract these singularities, obtain a regular
integrand, expand the integrand in $\eps$ and then finally integrate the
integrand numerically~\cite{Groote:1998wy}.

Let us describe our procedure in more detail. We write
\begin{eqnarray}\label{V4main}
\tilde\Pi_4(0)&=&{\cal N}_4\int D(x,m)^5d^Dx
  ={\cal N}_4\int D(x,m)(D(x,m)-\Delta(x)+\Delta(x))^4d^Dx\nonumber\\
  &=&{\cal N}_4\int D(x,m)\Big((D(x,m)-\Delta(x))^4+4(D(x,m)-\Delta(x))^3
  \Delta(x)\nonumber\\&&
  +6\left(D(x,m)-\Delta(x)\right)^2\Delta(x)^2
  +4\left(D(x,m)-\Delta(x)\right)\Delta(x)^3+\Delta(x)^4\Big)d^Dx
\end{eqnarray}
with
\begin{equation}
\Delta(x)=D(x,0)+D_1(x,0)+D_2(x,0)
\end{equation}
where the functions $D(x,0)$, $D_1(x,0)$, $D_2(x,0)$ subtract singularities of
$D(x,m)$ at small $x$. Their explicit expressions are given in Appendix~B. In
fact, these functions represent a formal expansion of the massive propagator
$D(x,m)$ in the mass parameter $m$ for small $m$ since the real expansion
parameter is the dimensionless quantity $mx$.

One propagator factor $D(x,m)$ in the integrand is left unsubtracted. At large
$x$ it provides an IR cutoff. The last two terms of the integrand in
Eq.~(\ref{V4main}) can be integrated analytically. They contain all poles in
$\eps$ and are expressible through Euler's $\Gamma$-functions. As expected,
the pole part coincides with the expression in Eq.~(\ref{poleparts}). Because
the analytical expression (as given up to order $\eps^3$ in Appendix~C) is
rather lengthy, we present only its numerical evaluation,
\begin{eqnarray}
\tilde\Pi_4^{\rm ana}(0)&=&m^6\Big(-2.5\eps^{-4}- 11.6666667\eps^{-3}
  -31.701389\eps^{-2}-67.528935\eps^{-1}\nonumber\\&&
  -15871.965743-142923.10240\eps-701868.64762\eps^2-2486982.5547\eps^3
  +O(\eps^4)\Big).\qquad
\end{eqnarray}

\begin{figure}[t]\begin{center}
\epsfig{figure=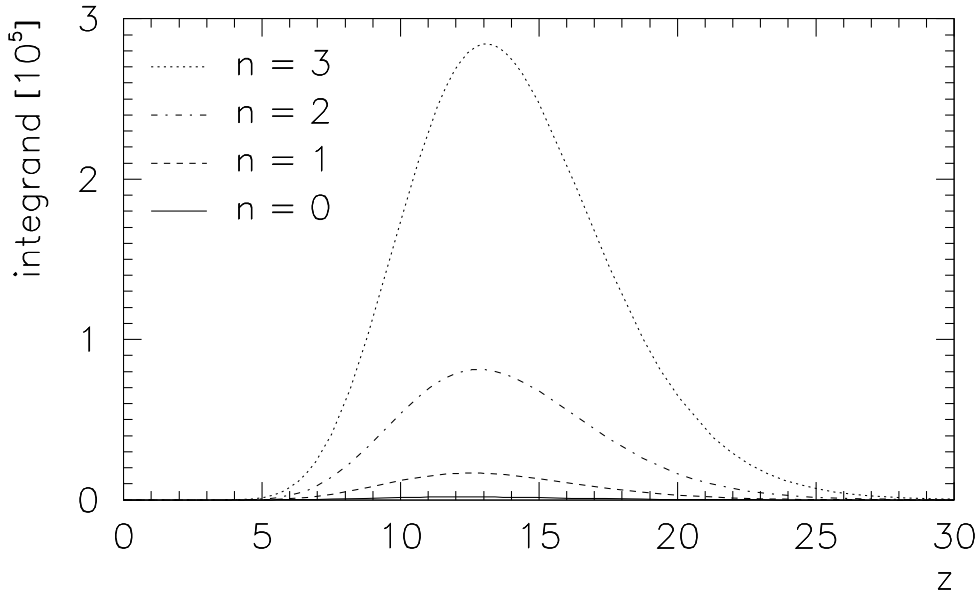, scale=0.8}
\caption{\label{fig3}Integrands for numerical integration in case of the
  four-loop bubble at different orders in $\eps$}
\end{center}\end{figure}

The first three terms in Eq.~(\ref{V4main}) can be integrated numerically for
$D=4$ (i.e.\ no regularization is necessary) since they are regular at small
$x$. The analytical expression for the functions to be integrated is rather
long. The zeroth order $\eps$-coefficient is found in Appendix~D (see
also the discussion of the integration procedure given in Appendix~D). However,
as shown in Fig.~\ref{fig3}, the functions themselves show a very smooth
behaviour which renders the numerical integration rather simple. We obtain
\begin{equation}
\tilde\Pi_4^{\rm num}(0)=m^6\Big(15731.745122+142349.56687\eps
  +699112.42072\eps^2+2468742.6339\eps^3+O(\eps^4)\Big).
\end{equation}
The sum of both parts gives
\begin{eqnarray}
\tilde\Pi_4(0)&=&m^6\Big(-2.5\eps^{-4}-11.6666667\eps^{-3}
  -31.701389\eps^{-2}-67.528935\eps^{-1}\nonumber\\&&
  -140.220621-573.53553\eps-2756.22690\eps^2-18239.9208\eps^3+O(\eps^4)
  \Big)\qquad
\end{eqnarray}
which confirms the known result~\cite{Laporta:2002pg}. The difference to the
results of~\cite{Laporta:2002pg} is within the accuracy of the numerical
evaluation of the integrals. Therefore, we provide an independent check of
this important quantity in full agreement with the results of Laporta.

It is not difficult to extend the analysis to higher orders in $\eps$ or to a
larger number of significant digits in the coefficients of the
$\eps$-expansion. However, since the technique is rather straightforward and
simple we do not consider it worthwhile to extend the calculations into these
directions. If the need arises, the potential user can tailor and optimize his
or her programming code to obtain any desired accuracy and/or order of the
$\eps$-expansion. In our evaluation we have used standard tools provided by
the MATHEMATICA package which allows reliably control the accuracy of numerical
calculations. Even at this early step of improvement it is obvious that our
algorithm is extremely simple and efficient.

\subsection{Five-loop vacuum bubble}
In this subsection we present results for the next order $p^2=0$
sunrise-type diagram, the five-loop vacuum bubble. We have chosen to extend
our calculation to the five-loop case since there exist no results on the
five-loop bubble in the literature. The integral representation of the
five-loop bubble is given by 
\begin{equation}
\tilde\Pi_5(p^2=0)={\cal N}_5\int D(x,m)^6d^Dx.
\end{equation}
Evaluating numerically the analytical part one obtains
\begin{eqnarray}
\tilde\Pi_5^{\rm ana}(0)&=&m^8\Big(3\eps^{-5}+16.5\eps^{-4}
  +51.95833\eps^{-3}+125.6715\eps^{-2}+259.9876\eps^{-1}\nonumber\\&&
  -1360392.5934-16888723.177\eps-111392297.46\eps^2
  -518606741.1\eps^3+O(\eps^4)\Big)\qquad
\end{eqnarray}
while the numerical integration of the nonsingular part gives
\begin{eqnarray}
\tilde\Pi_5^{\rm num}(0)&=&m^8\Big(1360739.9485+16886269.683\eps\nonumber\\&&
  +111360751.91\eps^2+518295438.0\eps^3+O(\eps^4)\Big).
\end{eqnarray}
The sum of both contributions is given by
\begin{eqnarray}
\tilde\Pi_5(0)&=&m^8\Big(3\eps^{-5}+16.5\eps^{-4}+51.95833\eps^{-3}
  +125.6715\eps^{-2}+259.9876\eps^{-1}\nonumber\\&&
  +347.3551-2453.494\eps-31545.55\eps^2-311303.1\eps^3+O(\eps^4)\Big)\qquad.
\end{eqnarray}
One observes huge cancellation effects between the terms obtained by the
analytical calculation and the numerical integration. Apparently the
subtraction procedure chosen here is non-optimal. As mentioned before, the
subtraction procedure should really be optimized for any given problem in
order to avoid a necessity to retain high numerical precision at intermediate
steps of calculations. Nevertheless, our nonoptimized simple subtraction
procedure already works quite reliably with available standard computational
tools.

In this section we have described how the configuration space technique works 
for the case of a trivial numerator. However, our method is applicable and
quite efficient also for nontrivial numerator factors as shown in the next
section.

\section{Irreducible numerator factors in sunrise-type diagrams}
For vacuum bubbles of a given topology there can be more than one master
integral. For example, in the case of the four-loop sunrise-type bubble
topology Laporta identified a second master integral $V_2$ which has a
nontrivial numerator factor which cannot be further
reduced~\cite{Laporta:2002pg}. In the momentum representation the non-trivial
numerator factors contain scalar products of loop momenta which cannot be
further canceled against denominator factors. To be precise, in the case of a
$n$-loop vacuum bubble with $(n+1)$ massive lines, the numerator factor is
trivially reducible for $n<3$. However, starting with $n=3$ a numerator
factor is no longer directly reducible in the general case. 

\subsection{Three-loop vacuum bubble with irreducible numerator}
As an example let us consider the case of a numerator $(k_i\cdot k_j)$ for a
vacuum bubble with four massive lines. In momentum space the integral of
interest reads
\begin{equation}
\tilde\Pi_3^*(0)=\left(\pi^{D/2}\Gamma(3-D/2)\right)^{-3}
  \int\frac{(k_1\cdot k_3)d^Dk_1d^Dk_2d^Dk_3}{(m_1^2+k_1^2)
  (m_2^2+(k_2-k_1)^2)(m_3^2+(k_3-k_2)^2)(m_4^2+k_3^2)}.
\end{equation}
An expansion of the numerator in terms of invariants,
\begin{equation}
2(k_1\cdot k_3)=k_1^2+k_3^2-(k_1-k_3)^2
\end{equation}
contains a structure $(k_1-k_3)^2$ which is absent in the denominator.
Therefore, the numerator is not directly (trivially) reducible to unity.

In configuration space it is not difficult to treat such a non-trivial
numerator since one can reduce the numerator to derivatives of the
propagators. The main features of the reduction method are more easily
discussed in terms of the equal mass case. The generalization to nonequal mass
case is evident.  In the above example one can write (using $m_1=m_2=\ldots=m$)
\begin{equation}
\tilde\Pi_3^*(0)={\cal N}_3\int D(x,m)^2\left(\partial_\mu D(x,m)\right)
  \left(\partial^\mu D(x,m)\right)d^Dx
\end{equation}
The integration-by-parts identity takes the form
\begin{equation}\label{ibp1}
\int\partial_\mu\left(D(x,m)\cdots D(x,m)\right)d^Dx=0
\end{equation}
which, as usual, should be used with the necessary caution.\footnote{Why this
caution is necessary is illustrated by the simple example of a massless
propagator. The massless propagator satisfies the equation
$-\partial^2D(x,0)=\delta(x)$ where $\partial^2=\partial^\mu\partial_\mu$.
Integrating the equation over the whole space and dropping the total
derivative on the left hand side one obtains the contradiction $0=1$
(for details cf.\ \cite{Isaev}).}
A further useful identity is given by
\begin{equation}\label{ibp2}
\int \partial^2\left(D(x,m)\cdots D(x,m)\right)d^Dx=0.
\end{equation}
The identities in Eq.~(\ref{ibp1}) and~(\ref{ibp2}) lead to a relation between
integrals with derivatives. In our example involving four propagators one
obtains
\begin{equation}
\int D(x,m)^2\left(\partial_\mu D(x,m)\left(\partial^\mu D(x,m)\right)\right)
  d^Dx=-\frac13\int D(x,m)^3\partial^2D(x,m)d^Dx.
\end{equation}
For the last integral we use $(-\partial^2+m^2)D(x,m)=\delta(x)$ to end up
with
\begin{equation}
\int D(x,m)^3\partial^2D(x,m)d^Dx=m^2\int D(x,m)^4d^Dx-D(0,m)^3.
\end{equation}
The value of $D(0,m)$ can be found by integration in momentum
space,\footnote{$D(0,m)$ can also be obtained by taking the limit $x\to 0$ on
the right hand side of Eq.~(\ref{eqn02}).}
\begin{equation}
D(0,m)=\frac1{(2\pi)^D}\int\frac{d^Dp}{m^2+p^2}
  =\frac{2\pi^{D/2}}{(2\pi)^D\Gamma(D/2)}\int_0^\infty\frac{p^{D-1}dp}{p^2+m^2}
  =\frac{m^{D-2}}{(4\pi)^{D/2}}\Gamma(1-D/2).
\end{equation}
Finally we have 
\begin{equation}
\tilde\Pi_3^*(0)=-\frac{m^2}3\tilde\Pi_3(0)+\frac13{\cal N}_3D(0,m)^3
\end{equation}
where $\tilde\Pi_3(0)$ is a three-loop sunrise-type bubble diagram without
numerators, as calculated earlier. This relation can explicitly be checked.

\subsection{Four-loop vacuum bubble with irreducible numerator}
As a more realistic example we consider a four-loop diagram that appears as a
second independent master integral $V_2$ of the sunrise topology in the
classification of Ref.~\cite{Laporta:2002pg}. In momentum space the second
master integral $V_2$ has the additional numerator factor $(k_1\cdot k_4)^2$
as compared to the first master integral $V_1$ with unity in the numerator.
The second master integral reads
\begin{eqnarray}
\tilde\Pi_4^*(0)&=&\left(\pi^{D/2}\Gamma(3-D/2)\right)^{-4}\ \times\\&&
  \int\frac{(k_1\cdot k_4)^2d^Dk_1d^Dk_2d^Dk_3d^Dk_4}{(m_1^2+k_1^2)
  (m_2^2+(k_2-k_1)^2)(m_3^2+(k_3-k_2)^2)(m_4^2+(k_4-k_3)^2)(m_5^2+k_4^2)}.
  \nonumber
\end{eqnarray}
Turning again to the equal mass case and using the configuration space
representation, this integral can be written as
\begin{equation}
\tilde\Pi_4^*(0)={\cal N}_4\int D(x,m)^3
  (\partial_\mu\partial_\nu D(x,m))(\partial_\mu\partial_\nu D(x,m))d^Dx.
\end{equation}
It is apparent that by using integration-by-parts techniques this integral
cannot be reduced to scalar integrals and/or integrals containing
d'Alembertians. The easiest way to evaluate such an integral is to compute the
derivatives directly. This is done by using
\begin{equation}
\frac1z\frac{d}{dz}\left(z^{-\nu}K_\nu(z)\right)
  =-\left(z^{-\nu-1}K_{\nu+1}(z)\right).
\end{equation}
This relation can be iterated and gives results for arbitrary high order
derivatives of Bessel functions $K_\lambda(z)$ in terms of the same class of
Bessel functions with shifted indices and powers in $z$. For the first
derivative we obtain
\begin{equation}
\partial_\mu D(x,m)=-x_\mu\frac{m^{2\lambda+2}}{(2\pi)^{\lambda+1}}
\frac{K_{\lambda+1}(mx)}{(mx)^{\lambda+1}}.
\end{equation}
Since the resulting analytical expression for the given line of the diagram
lies in the same class as the original line, the procedure of evaluation of the
integral is similar to the usual one. However, we cannot use the second
derivative
\begin{equation}
\label{secondDer}
\partial_\mu\partial_\nu D(x,m)=\frac{m^{2\lambda+2}}{(2\pi)^{\lambda+1}
  (mx)^{\lambda+1}}\left(g_{\mu\nu}K_{\lambda+2}(mx)-\frac{x_\mu x_\nu}{x^2}
  K_{\lambda+1}(mx)\right)
\end{equation}
directly under the integration sign. The reason is that if the propagator is
considered as a distribution, there is a $\delta$-function singularity at the
origin which is not taken into account in Eq.~(\ref{secondDer}). Indeed,
contracting the indices $\mu$ an $\nu$ in Eq.~(\ref{secondDer}) one obtains
\begin{equation}
\label{Dalam}
\partial_\mu\partial_\mu D(x,m)=m^2D(x,m)
\end{equation}
while the correct equation for the propagator reads
$(-\partial^2+m^2)D(x,m)=\delta(x)$. Thus, the straightforward evaluation of
derivatives is always valid only for $x=0$. The behaviour at the origin
($x=0$) requires a special consideration. In practice, to treat this situation
one should not use higher order derivatives but stay at the level of first
derivatives.

In order to deal with this situation we introduce another master integral
\begin{equation}
\tilde\Pi_4^{**}(0)={\cal N}_4\int D(x,m)\partial_\mu D(x,m)\partial_\nu
  D(x,m)\partial^\mu D(x,m)\partial^\nu D(x,m)d^Dx.
\end{equation}
The relation between the two master integrals $\Pi_4^*(0)$ and $\Pi_4^{**}(0)$
is found to be
\begin{equation}\label{v2v2bar}
\tilde\Pi_4^*(0)=3\tilde\Pi_4^{**}(0)-\frac18m^4\tilde\Pi_4(0)
  -\frac78m^2{\cal N}_4D(0,m)^4.
\end{equation}
The quantity $\tilde\Pi_4^{**}(0)$ can be calculated with the explicit
use of first order derivatives within our technique. We obtain the analytical
result for the pole part
\begin{equation}
\tilde\Pi_4^{**}(0)=m^{10}\left(-\frac3{8\eps^4}-\frac{277}{144\eps^3}
  -\frac{37837}{6912\eps^2}-\frac{4936643}{414720\eps}+O(\eps^0)\right)
\end{equation}
and the $\eps$-expansion in the form 
\begin{eqnarray}
\tilde\Pi_4^{**}(0)
  &=&m^{10}\Big(-0.375\eps^{-4}-1.923611\eps^{-3}-5.474103\eps^{-2}
  -11.90356\eps^{-1}\nonumber\\&&
  -27.99303-104.5384\eps-663.6123\eps^2-3703.241\eps^3+O(\eps^4)\Big).
\end{eqnarray}
Since the results for $\tilde\Pi_4(0)$, $\tilde\Pi_4^{**}(0)$, and $D(0,m)^4$
are known, Eq.~(\ref{v2v2bar}) can be used to obtain the final result for the
original integral,
\begin{eqnarray}
\tilde\Pi_4^*(0)&=&m^{10}\Big(-1.6875\eps^{-4}-7.8125\eps^{-3}
  -21.20964\eps^{-2}-44.76955\eps^{-1}\nonumber\\&&
  -97.07652-290.9234\eps-1719.809\eps^2-8934.731\eps^3
  +O(\eps^4)\Big)\qquad
\end{eqnarray}
which again verifies the result given in Ref.~\cite{Laporta:2002pg}.

Differentiation of the massive propagator leads to expressions of a similar
functional form which makes the configuration space technique a universal tool
for calculating any master integral of the sunrise topology. This technique is
also useful for finding master integrals. Indeed, new master integrals appear
when there is a possibility to add new derivatives into integrands which
cannot be eventually removed by using the equations of motion or
integration-by-parts recurrence relations. But once again: without explicit
inclusion of the $\delta$-function only one derivative is allowed. Otherwise
one runs into problems not seeing some parts (tadpoles) of the result.
Therefore, the new master integral should contain just one derivative for each
line excepting one line. For instance, in the five-loop case (six propagators)
there will be only two master integrals.

\section{Summary and conclusions}
We have suggested a new efficient technique to compute diagrams of the 
sunrise-type topology with any number of loops at any order of the 
$\eps$-expansion for any mass configuration. For a given number of loops the
sunset-type topology constitutes only a small fraction of the whole set of
multi-loop topologies that need to be calculated. Nevertheless, our results on
the subset of sunrise-type diagrams provide a necessary check on multi-loop
results calculated by other techniques and therefore can be very useful for
many multi-loop calculations. We have also worked out a few examples with
nontrivial numerator factors. Valuable by itself, our method can be used to
check the results of other techniques.

\subsection*{Acknowledgement}
We thank K.G.~Chetyrkin for his constructive, yet exacting criticism in the
course of our work on the application of $x$-space techniques to the
evaluation of sunrise diagrams. We thank R. Delbourgo for his kind attention
and enthusiasm in advertising and developing $x$-space techniques and
G.~Passarino for a communication. We would also like to thank J.~Gasser for 
his interest and discussions. AAP thanks V.A.~Matveev for encouragement,
attention and support. This work is partially supported by the Volkswagen
grant I/77~788, the RFBR grants \# 02-01-00601 and \# 03-02-17177, and the
Estonian target financed project No.~0182647s04. S.~Groote acknowledges
support from a grant given by the Graduiertenkolleg ``Eichtheorien'', Mainz
University.

\begin{appendix}

\section{Singular contributions for arbitrary masses}
In the following we present complete results for the singular parts of
sunrise-type diagrams with arbitrary masses up to four-loop order. The results
are given in the $\overline{\rm MS}$-scheme in the Euclidean domain.
\begin{eqnarray}
\lefteqn{\tilde\Pi_1^s(p,m_1,m_2)\ =\ \frac1\eps,}\nonumber\\[7pt]
\lefteqn{\tilde\Pi_2^s(p,m_1,m_2,m_3)\ =\ -\frac1{2\eps^2}\sum_im_i^2
  -\frac1{4\eps}\left(p^2+2\sum_im_i^2(3-2\ell_i)\right),}\nonumber\\[7pt]
\lefteqn{\tilde\Pi_3^s(p,m_1,m_2,m_3,m_4)\ =\ \frac1{6\eps^3}\sum_{i\neq j}
  m_i^2m_j^2
  +\frac1{12\eps^2}\left(p^2\sum_im_i^2-\sum_im_i^4+2\sum_{i\neq j}m_i^2m_j^2
  \left(4-3\ell_i\right)\right)\,+}\nonumber\\&&\kern-8pt
  +\frac1{72\eps}\left(2p^4+9p^2\sum_im_i^2(3-2\ell_i)-9\sum_im_i^4(5-2\ell_i)
  +6\sum_{i\neq j}m_i^2m_j^2(20-24\ell_i+3\ell_i^2+6\ell_i\ell_j)\right),
  \nonumber\\[7pt]
\lefteqn{\tilde\Pi_4^s(p,m_1,m_2,m_3,m_4,m_5)\ =\ -\frac1{24\eps^4}
  \sum_{i\neq j\neq k}m_i^2m_j^2m_k^2\,+}\nonumber\\&&
  -\frac1{48\eps^3}\left(p^2\sum_{i\neq j}m_i^2m_j^2
  -\sum_{i\neq j}(m_i^4m_j^2+m_i^2m_j^4)
  +2\sum_{i\neq j\neq k}m_i^2m_j^2m_k^2(5-4\ell_i)\right)\,+\nonumber\\&&
  -\frac1{288\eps^2}\Bigg(2p^4\sum_im_i^2-6p^2\sum_im_i^4+2\sum_im_i^6
  +3p^2\sum_{i\neq j}m_i^2m_j^2(11-8\ell_i)\,+\nonumber\\&&\qquad
  -6\sum_{i\neq j}(m_i^4m_j^2+m_i^2m_j^4)(7-4\ell_i)
  +12\sum_{i\neq j\neq k}m_i^2m_j^2m_k^2(15-20\ell_i+2\ell_i^2+6\ell_i\ell_j)
  \Bigg)
  \,+\nonumber\\&&
  -\frac1{1728\eps}\Bigg(3p^6+2p^4\sum_im_i^2(35-24\ell_i)
  -18p^2\sum_im_i^4(21-8\ell_i)+2\sum_im_i^6(77-24\ell_i)\,+\nonumber\\&&\qquad
  +9p^2\sum_{i\neq j}m_i^2m_j^2(71-88\ell_i+8\ell_i^2+24\ell_i\ell_j)
  -216\sum_{i \neq j}(m_i^4m_j^2-m_i^2m_j^4)\ell_i\,+\nonumber\\&&\qquad
  -18\sum_{i\neq j}(m_i^4m_j^2+m_i^2m_j^4)
  (49-56\ell_i+4\ell_i^2+12\ell_i\ell_j)\,+\nonumber\\&&\qquad
  +24\sum_{i\neq j\neq k}m_i^2m_j^2m_k^2(105-180\ell_i+30\ell_i^2
  +90\ell_i\ell_j-2\ell_i^3-18\ell_i^2\ell_j-12\ell_i\ell_j\ell_k)\Bigg)
\end{eqnarray}
where $\ell_i=\ln(m_i^2/\mu^2)$. The indices $i$, $j$, and $k$ run over all
mass indices. One can check that the general results listed in this Appendix
reproduce the equal mass results listed in the main text.

\section{Explicit form of subtraction terms for the small $x$ singularities}
The leading singularity at small $x$ is given by the massless
propagator of the form
\begin{equation}
D(x,0)=\frac{\Gamma(\lambda)}{4\pi^{\lambda+1}x^{2\lambda}}.
\end{equation}
The next order of the small $x$-expansion for the propagator $D(x,m)$ 
is explicitly given by
\begin{equation}
D_1(x,0)=\frac1{4\pi^{\lambda+1}x^{2\lambda}}\left(\pfrac x2^2
  \frac{\Gamma(\lambda)}{1-\lambda}-\pfrac x2^{2\lambda}
  \frac{\Gamma(1-\lambda)}\lambda\right).
\end{equation}
This term is suppressed relative to the first term by one power of $x^2$ at
small $x$ in four-dimensional space-time (however, this is not the case for
two-dimensional space-time with $\lambda=0$). The term 
\begin{equation}
D_2(x,0)=\frac1{4\pi^{\lambda+1}x^{2\lambda}}\pfrac x2^2
\left(\pfrac x2^2\frac{\Gamma(\lambda)}{2(1-\lambda)(2-\lambda)}
  -\pfrac x2^{2\lambda}\frac{\Gamma(1-\lambda)}{\lambda(\lambda+1)}\right)
\end{equation}
is further suppressed by one power of $x^2$ at small $x$. Therefore, the full
subtraction of the three terms gives a rather smooth behaviour at small $x$
which is sufficient to obtain a regular integrand for the numerical
integration. 

\newpage

\section{Analytical results for the four-loop sunrise diagram}
In this appendix we present some more details of our calculations for the
four-loop sunrise diagram. For the analytical evaluation we take the last two
terms of the integrand from Eq.~(\ref{V4main}). One has to integrate a product
of two Bessel functions with powers of $x$ which can be done analytically.
The explicit expression for the $\eps$-expansion of that part of the integral
which is evaluated analytically reads
\begin{eqnarray}
\tilde\Pi_4^{\rm ana}(0)&=&m^6\Bigg\{-\frac5{2\eps^4}-\frac{35}{3\eps^3}
  -\frac{4565}{144\eps^2}-\frac{58345}{864\eps}\nonumber\\&&
  -\frac{1456940638037}{7779240000}-\frac{17099\pi^2}{24}
  -\frac{3857\pi^4}{10}+\frac{2525968\zeta(3)}{105}\nonumber\\&&
  +\Bigg(-\frac{55171475321621447}{1633640400000}+\frac{2457509\pi^2}{144}
  -\frac{1292537\pi^4}{175}\nonumber\\&&\qquad
  +\frac{6752474831\zeta(3)}{44100}+16530\pi^2\zeta(3)+59508\zeta(5)\Bigg)\eps
  \nonumber\\&&
  +\Bigg(-\frac{10610679621089130529}{68612896800000}
  +\frac{92781949\pi^2}{864}-\frac{4290113759\pi^4}{110250}
  -\frac{22591\pi^6}{14}\nonumber\\&&\qquad
  +\frac{952412727629\zeta(3)}{9261000}+244476\pi^2\zeta(3)-168606\zeta(3)^2
  +\frac{32210272\zeta(5)}{35}\Bigg)\eps^2\nonumber\\&&
  +\Bigg(-\frac{5963907632629558995931}{14408708328000000}
  +\frac{1325204033\pi^2}{5184}\nonumber\\&&\qquad
  -\frac{464379085699\pi^4}{6615000}-\frac{48529231\pi^6}{2205}
  \nonumber\\&&\qquad
  -\frac{312138383154103\zeta(3)}{277830000}+\frac{7285043\pi^2\zeta(3)}{6}
  +43529\pi^4\zeta(3)-\frac{238229084\zeta(3)^2}{105}\nonumber\\&&\qquad
  +\frac{13583011297\zeta(5)}{2940}+247950\pi^2\zeta(5)+1190160\zeta(7)\Bigg)
  \eps^3+O(\eps)^4\Bigg\}.
\end{eqnarray}
This expression shows the real complexity of the calculation which reveals
itself in the structure of the results. The main feature is that the terms
cannot be simultaneously simplified to all orders in $\eps$. By a special
choice of the normalization factor one can make the leading term and, in fact,
even all pole terms simple, but then the higher order terms contain rather
lengthy combinations of transcendental numbers that are not reducible in terms
of standard quantities such as the Riemann $\zeta$-functions. Note also that
the rational coefficients of transcendental numbers are very big and there is a
huge numerical cancellation between the rational and transcendental parts 
of the answer~(see also the discussion in Ref.~\cite{Groote:1999cx}).

\section{Integrand for numerical integration}
For the numerical evaluation we take the first three terms of the integrand
from Eq.~(\ref{V4main}). One has to integrate them numerically as there is a
product of three or more Bessel functions which is too complicated to be done
analytically. To find the $\eps$-expansion of the integral one has to first
expand the integrand in $\eps$. The expression for the $\eps$-expansion is
quite lengthy. We therefore give explicit results only for $\eps=0$. For this
part the integrand for the numerical integration over $z=mx$ reads
\begin{eqnarray}\label{integrand}
\Pi_4^{\rm num}(x)&=&m^6\Bigg(66-108l-144l^2+192l^3+\frac{384}{z^6}
  -\frac{384}{z^4}+\frac{768l}{z^4}+\frac{24}{z^2}-\frac{480l}{z^2}
  +\frac{576l^2}{z^2}\nonumber\\&&\qquad
  -\frac{111z^2}{16}+\frac{147lz^2}{2}-117l^2z^2+24l^3z^2+24l^4z^2+
  -\frac{165z^4}{32}\nonumber\\&&\qquad
  +\frac{201lz^4}{16}+\frac{9l^2z^4}{2}-24l^3z^4+12l^4z^4+\frac{75z^6}{512}
  -\frac{405lz^6}{128}+\frac{531l^2z^6}{64}\nonumber\\&&\qquad
  -\frac{15l^3z^6}{2}+\frac{9l^4z^6}{4}+\frac{375z^8}{2048}
  -\frac{825lz^8}{1024}+\frac{315l^2z^8}{256}-\frac{51l^3z^8}{64}
  +\frac{3l^4z^8}{16}\nonumber\\&&\qquad
  +\frac{1875z^{10}}{131072}-\frac{375lz^{10}}{8192}+\frac{225l^2z^{10}}{4096}
  -\frac{15l^3z^{10}}{512}+\frac{3l^4z^{10}}{512}\Bigg)K_1(z)\nonumber\\&&
  +m^6\Bigg(\frac{-512}{z^5}+\frac{384}{z^3}-\frac{768l}{z^3}+\frac{24}{z}
  +\frac{288l}{z}-\frac{384l^2}{z}-52z+120lz-64l^3z\nonumber\\&&\qquad
  -\frac{15z^3}{8}-21lz^3+48l^2z^3-24l^3z^3+\frac{75z^5}{32}
  -\frac{135lz^5}{16}+9l^2z^5-3l^3z^5\nonumber\\&&\qquad
  +\frac{125z^7}{512}-\frac{75lz^7}{128}+\frac{15l^2z^7}{32}
  -\frac{l^3z^7}{8}\Bigg)K_1(z)^2+m^6\frac{128K_1(z)^5}{z^2}.
\end{eqnarray}
Here $l=\ln(e^\gamma z/2)$, $z=mx$, and $\gamma=-\Gamma'(1)$ is Euler's
constant. As shown in Fig.~\ref{fig3}, the plot of this function as well as
the shapes of the corresponding functions in higher orders of $\eps$ are very
smooth and quite similar. The analytical expressions for higher orders in
$\eps$, however, become much longer. Note that the new functions $f_n(z)$
first appear at order $\eps^2$. 

The smoothness of the zeroth order integrand as shown in Eq.~(\ref{integrand})
implies that the numerical integration is quite easy to execute. Because
the integrand vanishes exponentially for large values of $z$ and has no
singularities of the kind $z\ln z$ for small values of $z$, the integration
can in principle range from $0$ to $\infty$. However, for practical reasons
we had to instruct MATHEMATICA (which we used for all of our calculations
presented here) that the integrand vanishes for $z=0$. On the other hand, the
asymptotic expansion of the integrand together with the integration measure
is dominated by the term
\begin{equation}
\frac{6\pi^2m^2}{512}z^{10}\ln^4(e^\gamma z/2)K_1(z)z^3\qquad
(K_\lambda(z)\to\sqrt{\frac\pi{2z}}e^{-z}\mbox{\ for\ }z\to\infty).
\end{equation}
Integrated over $z$ from $\Lambda$ to $\infty$, this part gives a contribution
\begin{equation}
\frac{6\pi^2m^2}{512}\Lambda^{25/2}\ln^4(e^\gamma\Lambda/2)e^{-\Lambda}
\end{equation}
and terms which are of subleading order. Therefore, the result can be
well estimated by
\begin{equation}
\tilde\Pi_4^{\rm num}(0)=2\pi^2\int_0^\infty\Pi_4^{\rm num}(x)x^3dx
  \approx 2\pi^2\int_0^\Lambda\Pi_4^{\rm num}(x)x^3dx
  +\frac{6\pi^2m^2}{512}\Lambda^{25/2}\ln^4(e^\gamma\Lambda/2)e^{-\Lambda}
\end{equation}
and $\Lambda$ can be adjusted in such a way that any desired precision is
obtained.

A possibility to avoid any kind of cutoff is to change the integration
variable in the sense that the interval $[0,\infty]$ is mapped onto $[0,1]$.
Then the integration can be done numerically with the additional information
that the integrand vanishes identically at both end points. Possible
transformations of this kind are for instance $z=\ln(1/t)$ or $z=(1-t)/t$
for $t\in[0,1]$.
 
\end{appendix}


\begin{thebibliography}{999}

\bibitem{SMrev}P.~Langacker (ed.),\\ ``Precision Tests of the Standard
  Electroweak Model'', World Scientific, 1995.

\bibitem{Hollik:1999uy}W.~Hollik,
``The electroweak Standard Model'',\\
  {\it Prepared for the ICTP Summer School in Particle Physics,
  Trieste, Italy, 21 Jun -- 9 Jul 1999}

\bibitem{Blumlein:qs}
J.~Bl\"umlein, F.~Jegerlehner, T.~Riemann, W.~Hollik and J.~H.~K\"uhn,
``Application Of Quantum Field Theory To Phenomenology --
  Loops And Legs In Quantum Field Theory'',
  Proceedings of the 6th International Symposium, Radcor 2002, and the
  6th Zeuthen Workshop, Kloster Banz, Germany, September 8 -- 13, 2002

\bibitem{threeloops}
J.G.~K\"orner, A.I.~Onishchenko, A.A.~Petrov and A.A.~Pivovarov,
  Phys.~Rev.~Lett.\ {\bf 91} (2003) 192002;
J.H.~K\"uhn, A.I.~Onishchenko, A.A.~Pivovarov and O.~L.~Veretin,
  Phys.~Rev.\ {\bf D68} (2003) 033018;
S.~Groote, J.G.~K\"orner, A.A.~Pivovarov,
  Eur.~Phys.~J.\ {\bf C24} (2002) 393

\bibitem{ibyparts}
  K.G.~Chetyrkin and F.V.~Tkachov,
  Nucl.~Phys.\ {\bf B192} (1981) 159;\\
  F.V.~Tkachov,
  Phys.~Lett.\ {\bf B100} (1981) 65

\bibitem{Broadhurst:1991fi}
D.J.~Broadhurst,
  Z.~Phys.\ {\bf C54} (1992) 599

\bibitem{Baikov:1996rk}P.A.~Baikov,
  Phys.~Lett.\ {\bf B385} (1996) 404

\bibitem{Chetyrkin:1996cf}K.G.~Chetyrkin, J.H.~K\"uhn and M.~Steinhauser,
  Nucl.~Phys.\ {\bf B482} (1996) 213

\bibitem{Laporta:1996mq}S.~Laporta and E.~Remiddi,
  Phys.~Lett.\ {\bf B379} (1996) 283

\bibitem{Tarasov:1997kx}O.V.~Tarasov,
  Nucl.~Phys.\ {\bf B502} (1997) 455

\bibitem{Gehrmann:1999as}T.~Gehrmann and E.~Remiddi,
  Nucl.~Phys.\ {\bf B580} (2000) 485

\bibitem{Melnikov:2000zc}K.~Melnikov and T.~van Ritbergen,
  Nucl.~Phys.\ {\bf B591} (2000) 515

\bibitem{Laporta:2001dd}
S.~Laporta,
  Int.~J.~Mod.~Phys.\ {\bf A15} (2000) 5087

\bibitem{Grozin:2002zb}A.G.~Grozin,
  Nucl.~Instrum.~Meth.\ {\bf A502} (2003) 610

\bibitem{Meijer}C.S.~Meijer, Proc.~Amsterdam~Akad.~Wet.\ {\bf 599} (1940) 702

\bibitem{ellRef}
B.~Almgren, Arkiv f\"or Fysik {\bf 38} (1967) 161;\\
S.~Bauberger, F.A.~Berends, M.~B\"ohm and M.~Buza,
  Nucl.~Phys.\ {\bf B434} (1995) 383

\bibitem{Mendels}E.~Mendels, Nuovo~Cim.\ {\bf 45 A} (1978) 87

\bibitem{Avdeev:1995eu}
L.V.~Avdeev,
  Comput.~Phys.~Commun.\ {\bf 98} (1996) 15

\bibitem{Groote:1999cn}
S.~Groote, J.G.~K\"orner and A.A.~Pivovarov,
  Phys.~Rev.\ {\bf D60} (1999) 061701

\bibitem{Schroder:2002re}Y.~Schr\"oder,
  Nucl.~Phys.~Proc.~Suppl.\ {\bf 116} (2003) 402

\bibitem{Larin:1986yt}
S.A.~Larin, V.A.~Matveev, A.A.~Ovchinnikov and A.A.~Pivovarov,
  Yad.~Fiz.\ {\bf 44} (1986) 1066;
I.I.~Balitsky, D.~Diakonov and A.V.~Yung,
  Phys.~Lett.\ {\bf B112} (1982) 71;
K.G.~Chetyrkin and S.~Narison,
  Phys.~Lett.\ {\bf B485} (2000) 145;
H.Y.~Jin, J.G.~K\"orner and T.G.~Steele,
  Phys.~Rev.\ {\bf D67} (2003) 014025

\bibitem{Sakai:1999qm}
T.~Sakai, K.~Shimizu and K.~Yazaki,
  Prog.~Theor.~Phys.~Suppl.\ {\bf 137} (2000) 121

\bibitem{Groote:2001vr}
S.~Groote and A.A.~Pivovarov,
  Eur.~Phys.~J.\ {\bf C21} (2001) 133;\\
  JETP Lett.\ {\bf 75} (2002) 221

\bibitem{bar}
A.A.~Ovchinnikov, A.A.~Pivovarov and L.R.~Surguladze,\\
  Sov.~J.~Nucl.~Phys.\ {\bf 48} (1988) 358;
  Int.~J.~Mod.~Phys.\ {\bf A6} (1991) 2025;
S.~Groote, J.G.~K\"orner and A.A.~Pivovarov,
  Phys.~Rev.\ {\bf D61} (2000) 071501 (2000);
``Analytical calculation of heavy baryon correlators 
  in NLO of perturbative QCD,''
  In {\it Batavia 2000, Advanced computing and analysis technique
  in physics research\/} 277-279 
[arXiv:hep-ph/0009218].

\bibitem{Narison:1994zt}
S.~Narison and A.A.~Pivovarov, 
  Phys.~Lett.\ {\bf B327} (1994) 341

\bibitem{effpot}
S.R.~Coleman and E.~Weinberg,
  Phys.~Rev.\ {\bf D7} (1973) 1888;
R.~Jackiw,
  Phys.~Rev.\ {\bf D9} (1974) 1686;
R.~Jackiw and S.~Templeton,
  Phys.~Rev.\ {\bf D23} (1981) 2291;
J.M.~Chung and B.K.~Chung,
  J.~Korean Phys.~Soc.\ {\bf 39} (2001) 971;
  Phys.~Rev.\ {\bf D59} (1999) 105014

\bibitem{Gross:1980br}
D.J.~Gross, R.D.~Pisarski and L.G.~Yaffe,
  Rev.~Mod.~Phys.\ {\bf 53} (1981) 43;\\
T.~Appelquist and R.D.~Pisarski,
  Phys.~Rev.\ {\bf D23} (1981) 2305;\\
T.~Hatsuda,
  Nucl.~Phys.\ {\bf A544} (1992) 27

\bibitem{Andersen:2000zn}
J.O.~Andersen, E.~Braaten and M.~Strickland,
  Phys.~Rev.\ {\bf D62} (2000) 045004

\bibitem{Nishikawa:2003js}
T.~Nishikawa, O.~Morimatsu and Y.~Hidaka,
  Phys.~Rev.\ {\bf D68} (2003) 076002

\bibitem{Rajantie:1996cw}
A.K.~Rajantie,
  Nucl.~Phys.\ {\bf B480} (1996) 729
[Erratum {\it ibid.\/} {\bf B513} (1998) 761]

\bibitem{nuclth}
L.~Platter, H.W.~Hammer and U.G.~Meissner,
  Nucl.~Phys.\ {\bf A714} (2003) 250;\\
J.F.~Yang, J.~Zhou and C.~Wu,
  Commun.~Theor.~Phys.\ {\bf 40} (2003) 461;\\
H.~Van Hees and J.~Knoll,
  Phys.~Rev.\ {\bf D65} (2002) 105005;\\
C.~Felline, N.P.~Mehta, J.~Piekarewicz and J.R.~Shepard,
  Phys.~Rev.\ {\bf C68} (2003) 034003

\bibitem{Witten:1979kh}
E.~Witten,
  Nucl.~Phys.\ {\bf B160} (1979) 57

\bibitem{Kajantie:2003ax}
K.~Kajantie, M.~Laine, K.~Rummukainen and Y.~Schr\"oder,
  JHEP {\bf 0304} (2003) 036

\bibitem{Post:1997dk}
P.~Post and K.~Schilcher,
  Phys.~Rev.~Lett.\ {\bf 79} (1997) 4088

\bibitem{Post:1996gg}
P.~Post and J.~B.~Tausk,
  Mod.~Phys.~Lett.\ {\bf A11} (1996) 2115

\bibitem{Gasser:1998qt}
J.~Gasser and M.E.~Sainio,
  Eur.~Phys.~J.\ {\bf C6} (1999) 297

\bibitem{Delbourgo:2003zi}
R.~Delbourgo and M.L.~Roberts,
  J.~Phys.\ {\bf A36} (2003) 1719

\bibitem{Bashir:2001ad}
A.~Bashir, R.~Delbourgo and M.L.~Roberts,
  J.~Math.~Phys.\ {\bf 42} (2001) 5553

\bibitem{Ligterink:1999mu}
N.E.~Ligterink,
  Phys.~Rev.\ {\bf D61} (2000) 105010;\\
B.~Kastening and H.~Kleinert,
  Phys.~Lett.\ {\bf A269} (2000) 50

\bibitem{Davydychev:1999ic}
A.I.~Davydychev and V.A.~Smirnov,
  Nucl.~Phys.\ {\bf B554} (1999) 391

\bibitem{Mendels:qe}E.~Mendels,
  J.~Math.~Phys.\ {\bf 43} (2002) 3011

\bibitem{Caffo:2002ch}
M.~Caffo, H.~Czy\.z and E.~Remiddi,
  Nucl.~Phys.\ {\bf B634} (2002) 309;\\
  Nucl.~Phys.~Proc.~Suppl.\ {\bf 116} (2003) 422

\bibitem{Bierenbaum:2003ud}
I.~Bierenbaum and S.~Weinzierl,
  Eur.~Phys.~J.\ {\bf C32} (2003) 67

\bibitem{Laporta:2002pg}S.~Laporta,
  Phys.~Lett.\ {\bf B549} (2002) 115

\bibitem{Passarino:2001wv}G.~Passarino,
  Nucl.~Phys.\ {\bf B619} (2001) 257

\bibitem{Groote:1998ic}
S.~Groote, J.G.~K\"orner and A.A.~Pivovarov,
  Phys.~Lett.\ {\bf B443} (1998) 269

\bibitem{Groote:1998wy}
S.~Groote, J.G.~K\"orner and A.A.~Pivovarov,
  Nucl.~Phys.\ {\bf B542} (1999) 515

\bibitem{Groote:2000kz}
S.~Groote and A.A.~Pivovarov,
  Nucl.~Phys.\ {\bf B580} (2000) 459

\bibitem{Watson}G.N.~Watson, 
``Theory of Bessel functions'', Cambridge, 1944

\bibitem{Chetyrkin:pr}
K.G.~Chetyrkin, A.L.~Kataev and F.V.~Tkachov,
  Nucl.~Phys.\ {\bf B174} (1980) 345

\bibitem{Terrano:1980af}A.E.~Terrano,
  Phys.~Lett.\ {\bf B93} (1980) 424

\bibitem{Bogoliubov}N.N.~Bogoliubov and D.V.~Shirkov,
  ``Quantum fields'', Benjamin, 1983

\bibitem{Prudnikov}A.P.~Prudnikov, Yu.A.~Brychkov and O.I.~Marichev,\\
  ``Integrals and Series'', Vol.~2, Gordon and Breach, New York, 1990

\bibitem{Gradshteyn}I.S.~Gradshteyn and I.M.~Ryzhik,\\
  ``Tables of integrals, series, and products'', Academic Press, 1994

\bibitem{Braaten:1995cm}
E.~Braaten and A.~Nieto,
  Phys.~Rev.\ {\bf D51} (1995) 6990

\bibitem{Groote:1999cx}
S.~Groote, J.G.~K\"orner and A.A.~Pivovarov,
  Eur.~Phys.~J.\ {\bf C11} (1999) 279

\bibitem{Davydychev:2000na}
A.I.~Davydychev and M.Y.~Kalmykov,
  Nucl.~Phys.\ {\bf B605} (2001) 266

\bibitem{Suzuki:2004ue}
A.T.~Suzuki and A.G.M.~Schmidt,\\
``$\varepsilon$-expansion for non-planar double-boxes'',
[arXiv:hep-ph/0401207]

\bibitem{Isaev}
A.P.~Isaev,
  Nucl.~Phys.\ {\bf B662} (2003) 461;\\
S.G.~Gorishnii and A.P.~Isaev,\\
  Theor.~Math.~Phys.\ {\bf 62} (1985) 232
  [Teor.~Mat.~Fiz.\ {\bf 62} (1985) 345]

\end{thebibliography}
\end{document}